# Demons and Abelian Projection QCD: Action and Crossover

Ken Yee

*Physics and Astronomy, L.S.U.*

and

*School of Law, Stanford University*

address: P.O. Box 9425, Stanford, CA 94309-9425

e-mail: kenton@leland.stanford.edu

March 2, 1995

## Abstract

$S_{APQCD}$, the Abelian projection QCD(APQCD) action, is evaluated using the demon method. For $SU(2)$, $S_{APQCD}$ at strong coupling is essentially the compact QED(CQED) action with $\beta_{CQED} = \frac{1}{2}\beta_{SU(2)}$; extended and higher representation plaquettes are absent. Since CQED deconfines when $\beta_{CQED} >\sim 1$, this relation must break down as $\beta_{SU(2)} \to 2$. Indeed we find $S_{APQCD}$ mutates: near $\beta_{SU(2)} \sim 2$ it gains additional operators, including an exogenous *negative* magnetic monopole mass shift. $S_{APQCD}$ for $SU(3)$ has similar behavior. The Appendix gives a brief explanation of the demon method.

A clear demonstration that monopole condensation is the origin of QCD confinement would be a notable achievement. To this end, 't Hooft [1, 2, 3] proposed that QCD monopoles are magnetic with respect to the $[U(1)]^{N-1}$ Cartan subgroup of color $SU(N)$. Full $SU(N)$ gauge symmetry obscures these charges and it is necessary to gauge fix at least the $SU(N)/[U(1)]^{N-1}$ symmetry to expose them. In this scenario monopoles are fixed-gauge manifestations of gauge field features responsible for QCD confinement. Only in special gauges does one have a picture of QCD confinement caused by monopole condensation. In other gauges the features causing confinement are still present but they do not look like magnetic monopoles [4].

Numerical studies have found that maximal Abelian(MA) gauge [5] is compelling for 't Hooft's hypothesis. Upon decomposing gauge field $A$ into purely diagonal($n$) and purely off-diagonal($ch$) parts

$$A = A^n + A^{ch}, \qquad (1)$$

the MA gauge condition $D^n_\mu A^{ch}_\mu \equiv \partial_\mu A^{ch}_\mu - ig[A^n_\mu, A^{ch}_\mu] = 0$ leaves a residual $[U(1)]^{N-1}$ gauge invariance under

$$\Omega_{\text{residual}} = \text{diag}(\exp^{-i\omega_1}, \cdots, \exp^{-i\omega_N}), \qquad \sum_{i=1}^N \omega_i = 0. \qquad (2)$$

Under $\Omega_{\text{residual}}$ the $N$ diagonal matrix elements $(A^n)_{ii}$ transform as neutral photon fields whereas the $N(N-1)$ offdiagonal matrix elements $(A^{ch})_{ij}$ transform as charged matter fields: $(A^n_\mu)_{ii} \to (A^n_\mu)_{ii} - \frac{1}{g}\partial_\mu \omega_i$ and, for $i \neq j$, $(A^{ch}_\mu)_{ij} \to (A^{ch}_\mu)_{ij} \exp^{-i(\omega_i - \omega_j)}$. Since $(A^{ch})_{ij}$ carries two different $U(1)$ charges, the $A^{ch}$ fields induce "interspecies" interactions between the $N$ photons. On the lattice the monopole currents are identified according to discretized versions [6] of $k_\mu \equiv \frac{1}{2\pi}\epsilon_{\mu\nu\lambda\delta}\partial_\nu f_{\lambda\delta}$ and $f_{\mu\nu} \equiv \partial_\mu A^n_\nu - \partial_\nu A^n_\mu$.

This procedure of using only the diagonal $A^n$ components of the $SU(N)$ gauge fields for measuring $k_\mu$ and $f_{\mu\nu}$ is called *Abelian projection*. Since $\text{tr} A^n_\mu = 0$ in $SU(N)$, an irreducible representation of $[U(1)]^{N-1}$ is

$$\theta^i_\mu \equiv (A^n_\mu)_{ii}. \qquad (3)$$

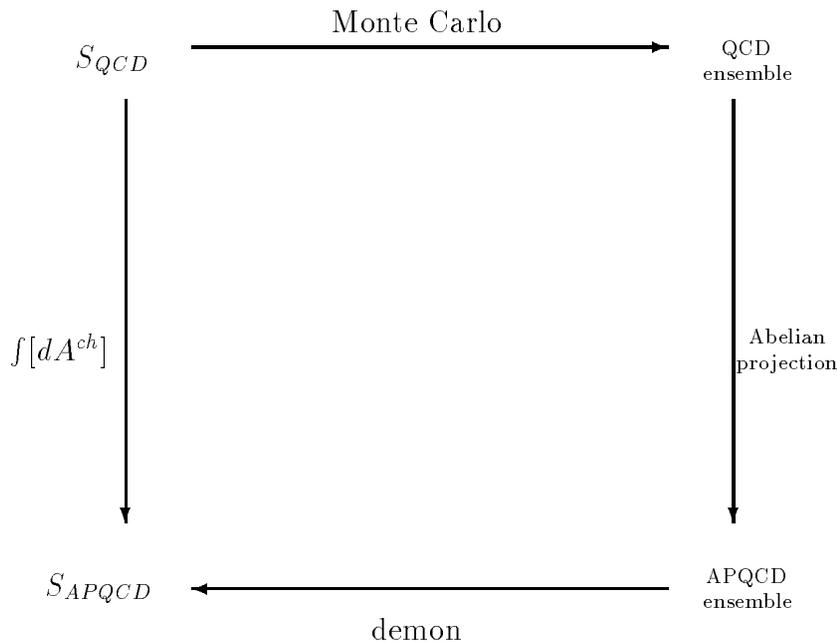

Figure 1: To integrate out $A^{ch}$ we: (i)generate an ensemble of importance sampling $SU(N)$ gauge configurations; (ii)project this ensemble to a $[U(1)]^{N-1}$ ensemble; and (iii)compute the $S_{APQCD}$ couplings using the microcanonical demon.

The $\theta^i$ transform as $\theta^i_\mu \to \theta^i_\mu - \frac{1}{g}\partial_\mu \omega_i$ and automatically obey constraint $\sum_{i=1}^{N} \theta^i_\mu = 0$. We shall refer to the quantum dynamics of the $N$ angles $\theta^i$ as Abelian projected QCD or APQCD. Equivalently, APQCD is the field theory obtained by integrating out $A^{ch}$ from QCD in MA gauge [7]. Its action $S_{APQCD}$ is formally defined as

$$-S_{APQCD}[\theta^1, \cdots, \theta^N] \equiv \log\left\{\int [dA^{ch}]\ \exp(-S_{QCD})\ \Delta_{FP}\ \delta[D^n_\mu A^{ch}_\mu]\right\} \quad (4)$$

where $\Delta_{FP}$ is the Faddeev-Popov determinant [8]. Monopoles arise in APQCD due to topological quantum fluctuations in the compact fields $\theta^i$.

While there is no guarantee that $S_{APQCD}$ has a simple form or is otherwise well-behaved, it is of central import due to *Abelian dominance* [9], the fact that $\theta^i$ Wilson loops in APQCD have predominantly the same string tension as $SU(N)$ Wilson loops in QCD. Abelian dominance has the following formal implication. If tr**W** is an $SU(N)$ Wilson loop and if $\langle \cdot \rangle_{QCD}$ and $\langle \cdot \rangle_{APQCD}$ refer respectively to $S_{QCD}$ and $S_{APQCD}$ expectation values, the APQCD operator $\mathcal{W}$ which obeys

$$\langle \mathcal{W} \rangle_{APQCD} = \text{tr} \langle \mathbf{W} \rangle_{QCD} \tag{5}$$

is

$$\mathcal{W} = \exp(+S_{APQCD}) \int [dA^{ch}] \, \exp(-S_{QCD}) \, \Delta_{FP} \, \delta[D_\mu^n A_\mu^{ch}] \, \text{tr}\mathbf{W}. \tag{6}$$

Abelian dominance means that $\mathcal{W}$, which in other gauges would be a complicated superposition of assorted $[U(1)]^{N-1}$-invariant operators of various sizes, shapes, and topologies, is (for string tension) well-approximated by a $\theta^i$ loop of the same size and shape as tr**W** in MA gauge. In other gauges the $\langle \mathcal{W} \rangle_{APQCD}$ string tension would be due to a *combination* of $S_{APQCD}$ effects *and* details (such as operator coefficients) of $\mathcal{W}$. In MA gauge, $S_{APQCD}$ *alone* determines string tension: given $S_{APQCD}$ one can reconstruct the QCD string tension using APQCD Wilson loops without reference to the RHS of (6). In this sense, in MA gauge $S_{APQCD}$ knows the QCD string tension.

Our numerical procedure for evaluating $S_{APQCD}$ is schematically summarized in Figure 1. First, focusing temporarily on $SU(2)$ we make an ensemble of importance sampling APQCD gauge configurations by applying the Abelian projection to a set of Monte Carlo $SU(2)$ gauge configurations at some coupling $\beta_{SU(2)}$. Then, seeking the action $S_{APQCD}$ which would reproduce this APQCD ensemble [7] in a Monte Carlo simulation, we state an ansatz for $S_{APQCD}$ and apply the microcanonical demon technique [10] to compute the parameters of this ansatz. This whole "inverse Monte Carlo" procedure is repeated starting from different QCD ensembles to determine how $S_{APQCD}$ varies with $\beta_{SU(2)}$.

The general $U(1)$-invariant action consistent with APQCD symmetries involves an infinity of operators. However, previous studies [7, 11] and, independently, the demon technique indicate that neither extended nor highly charged Wilson loops contribute substantially to $S_{APQCD}$. In particular, Table 1 shows the results of applying the demon to the ansatz

$$-S_A^{\text{ansatz}} \equiv \sum_{L=1}^{3} \sum_{q=1}^{3} \beta_q(L) \sum_{x,\mu<\nu} \cos q\Theta_{\mu\nu}(L) \qquad (7)$$

where $\cos q\Theta_{\mu\nu}(L)$ is an $L \times L$ plaquette in $U(1)$ representation $q$, given in terms of link angles $\theta_\mu^1$; $\beta_q(L)$ is its coupling the demon computes. Appendix A explains the technical idea underlying the demon technique. Figuratively, imagine a battalion of demons each carrying $M$ *coupled* thermometers whose temperatures correspond to the $M$ undetermined couplings in the ansatz action. ($M = 9$ for $S_A^{\text{ansatz}}$.) The demons thermalize with an APQCD configuration—the heat bath—by hopping from link to link. Each hopped-upon link is randomly updated, and the hopping demon absorbs or emits the corresponding energy($\equiv$action) decrement or increment created in the configuration by the update. The thermometers are coupled by requiring all their energies to simultaneously remain within a given range $[-E^0, E^0]$; if any proposed update violates this range it is rejected. Upon thermalization the couplings are read off by fitting the energy distributions of the demons' thermometers to $M$ Boltzmann distributions. Statistical errors are computed by jackknifing the demons. (The errors from jackknifing $SU(2)$ configurations are comparable.) In principle, if $S_A^{\text{ansatz}}$ contains all the operators of $S_{APQCD}$, the demon method tells us which couplings vanish and measures all the nonzero couplings exactly (modulo statistics). In practice, $S_A^{\text{ansatz}}$ is a truncated action which is unlikely to contain all $S_{APQCD}$ operators. Extensive numerical experiments with control ensembles (see Appendix and Ref. [15]) reveal that if important operators are missing the method yields readjusted "effective" coupling values. These effective values are not equal to the true values.

Table 1: $S_A^{\text{ansatz}}$ couplings in APQCD measured by the demon on a $14^3 \times 4$ lattice, where QCD is confining over the whole $\beta_{SU(2)}$ range given.

| parameter | $\beta_{SU(2)}$ | $L = 1$ | $L = 2$ | $L = 3$ |
|---|---|---|---|---|
| $\beta_1(L)$ | 1.0 | .50(.02) | .01(.01) | .02(.01) |
| $\beta_2(L)$ |  | -.02(.01) | .00(.00) | .01(.01) |
| $\beta_3(L)$ |  | .00(.00) | .00(.00) | .00(.01) |
| $\beta_1(L)$ | 2.0 | 1.03(.04) | .00(.00) | .02(.01) |
| $\beta_2(L)$ |  | .02(.00) | .00(.00) | .00(.01) |
| $\beta_3(L)$ |  | -.01(.01) | -.01(.01) | .00(.00) |
| $\beta_1(L)$ | 2.2 | 1.2(.03) | .00(.00) | .02(.01) |
| $\beta_2(L)$ |  | .04(.01) | .00(.00) | .00(.00) |
| $\beta_3(L)$ |  | -.01(.01) | -.03(.02) | -.02(.01) |

As illustrated in Table 1, $L = q = 1$ plaquettes dominate $S_{APQCD}$. Furthermore, there is no significant signal for any of the $L > 1$ plaquette couplings at any $\beta_{SU(2)}$. This result, unanticipated, has a substantial implication. Since BKT transformations of $L = 1$ actions do not lead to extended $L^3$ monopoles, this implies that, at least within our $\beta_{SU(2)}$ range, the fundamental dynamical degrees of freedom [11] for confinement in APQCD are pointlike $1^3$ rather than extended $L^3$ monopoles, as considered in [12]. In future work, it will be important to examine if this result survives the zero lattice spacing limit, e.g. if APQCD monopoles truly are pointlike.

On the other hand, the $L = 1$, $q = 2$ plaquette has a noticeable signal at $\beta_{SU(2)} = 2.2$ in Table 1. Therefore, we focus now on the $L = 1$ ansatz[1]

$$- S_B^{\text{ansatz}} = \sum_{q=1}^{3} \sum_{x, \mu < \nu} \beta_q \cos q\Theta_{\mu\nu} - \kappa \sum_{x,\mu} k_\mu(x) k_\mu(x). \qquad (8)$$

The $\kappa$ operator shifts the $q = 1$, $1^3$ monopole mass [12] implicit in $\beta_1 \cos \Theta_{\mu\nu}$, allowing the APQCD monopole mass to be independent of $\beta_1$. Of course, $\beta_q$ and $\kappa$ vary with $\beta_{SU(2)}$. Figure 2 shows $S_B^{\text{ansatz}}$ coefficients $\beta_1$, $\beta_2$ and $\kappa$, computed by the demon, as a function of $\beta_{SU(2)}$. $|\beta_3|$, not depicted, is always smaller than $|\beta_2|$, typically by a factor of $3 - 5$. Each $\beta_{SU(2)}$ configuration is generated fresh from a cold start so our data points do not contain any spurious correlations. Our $N_S^3 \times N_T = 20^3 \times 16$ lattices are all well inside the zero temperature phase; for the range of $\beta_{SU(2)}$ shown the APQCD Polyakov loop vanishes. As depicted, at strong coupling($\beta_{SU(2)} < 2$)

$$\beta_1 \sim \frac{1}{2}\beta_{SU(2)}, \quad \beta_{2,3} \sim 0, \quad \kappa \sim 0, \tag{9}$$

that is, $S_{APQCD}$ reduces to the compact QED(CQED) action at strong coupling. At weaker coupling($\beta_{SU(2)} > 2$) $\beta_2$ and $\kappa$ grow in magnitude but $\beta_1$ always remains the largest coupling.

---
[1] Unless otherwise specified, $L = 1$ is assumed in the remainder of this Note.

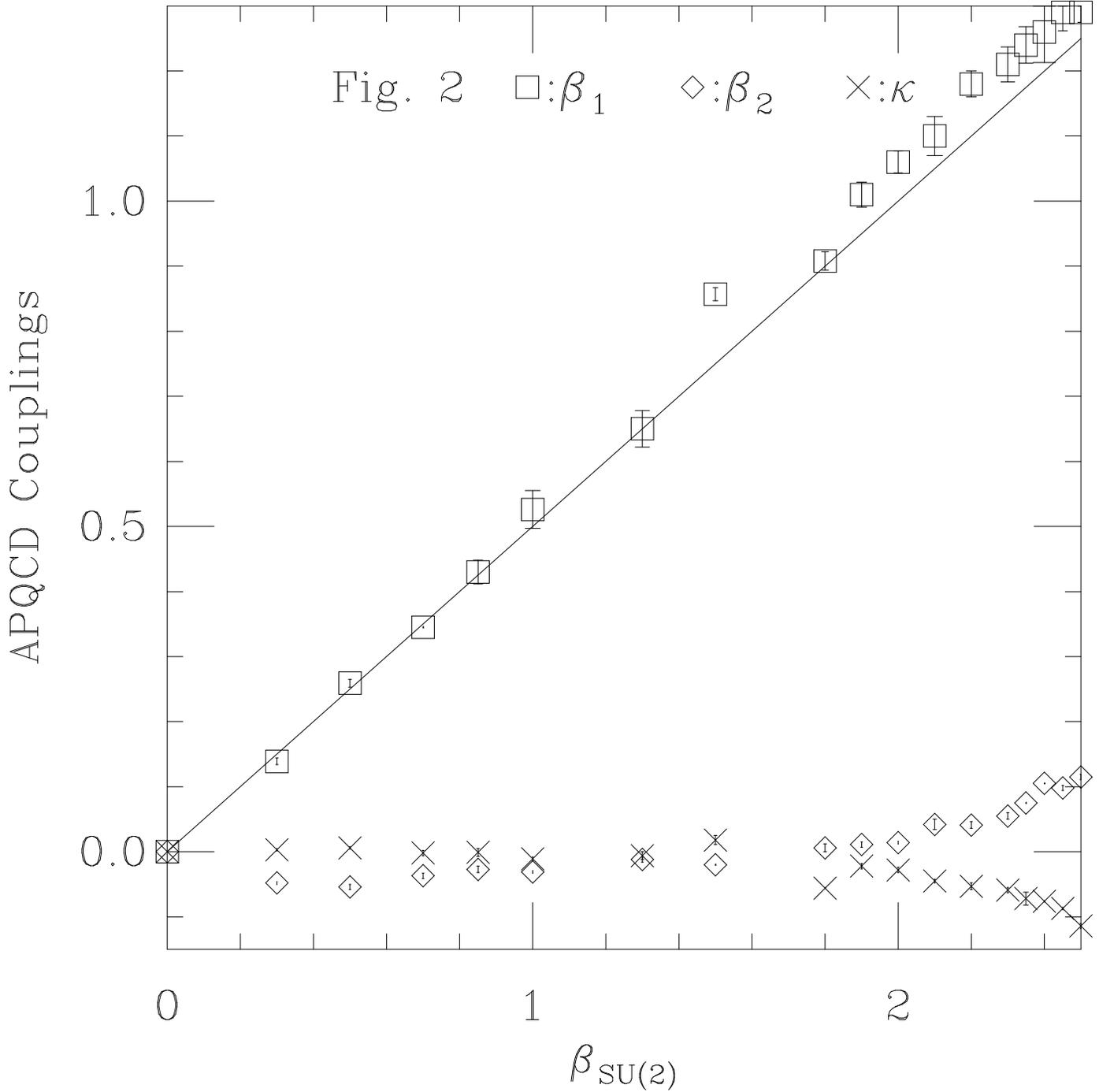

Figure 2: Figure 2 depicts $S_B^{\text{ansatz}}$ coefficients $\beta_1$, $\beta_2$ and $\kappa$ as a function of $\beta_{SU(2)}$. $|\beta_3|$, not depicted, is always smaller than $|\beta_2|$, typically by a factor of $3-5$. Our $20^3 \times 16$ lattices are all well inside the zero temperature phase for the $\beta_{SU(2)}$ range depicted. The bold $\beta_1 = \frac{1}{2}\beta_{SU(2)}$ line is a guide-to-eye.

Note that since monopoles are condensed when $\beta_{CQED} < 1$ in CQED [16], Figure 2 or Eq. (9) vicariously proves that $SU(2)$ monopoles are condensed when $\beta_{SU(2)} < 2$.

When $\beta_{SU(2)} > 2$, the situation is not so clear. In fact, Figure 2 suggests a paradox in the $\beta_{SU(2)} > 2$ region: how can APQCD maintain confinement in the continuum limit if CQED deconfines when $\beta_{CQED} > 1$? Clearly, either the meaning or validity of relation (9) must break down when $\beta_{SU(2)}$ is sufficiently large. Either (I)Abelian dominance does not survive the $\beta_{SU(2)} \sim 2$ crossover making $S_{APQCD}$ less pertinent at weaker coupling—see discussion pertaining to Eq. (6); or (II)$S_{APQCD}$ gains additional operators near $\beta_{SU(2)} \sim 2$; or a combination of (I) and (II). We do not have anything to say about (I) in this Note except that Abelian dominance apparently has been observed at all $\beta_{SU(2)} \leq 2.6$ [9].

(II) requires that when $\beta_{SU(2)} > 2$ $S_{APQCD}$ is no longer well described by $S_{CQED}$. Indeed, we can demonstrate this by simulating

$$-S_{CQED} = \sum_{x,\mu<\nu} \beta_{CQED} \cos \Theta_{\mu\nu} \Big|_{\beta_{CQED}=\beta_1(\beta_{SU(2)})} \tag{10}$$

(also on a $20^3 \times 16$ lattice) to see if it reproduces corresponding APQCD expectation values. As depicted in Figure 3, $S_{CQED}$ reproduces APQCD plaquette averages and monopole densities only in the $SU(2)$ strong coupling region. At weaker coupling the CQED simulations start to disagree dramatically with APQCD. This implies that at weaker coupling either other terms of $S_B^{\text{ansatz}}$ have become important or $S_B^{\text{ansatz}}$ itself is inadequate. In any case, this means $S_{APQCD}$ is not form-invariant between the strong and weak coupling regimes: at strong coupling $S_{APQCD}$ is well approximated by $S_{CQED}$; at crossover region $\beta_{SU(2)} \sim 2$ $S_{APQCD}$ mutates and develops substantial deviations from $S_{CQED}$. Inspection of Figure 2 reveals that a possible scenario[2] might be that $\kappa$, the exogenous magnetic monopole mass shift, becomes more and more *negative* at larger $\beta_{SU(2)}$. As negative monopole mass favors monopole condensation (compensating for a large $\beta_1$), the occurrence of a sufficiently negative $\kappa$ in $S_{APQCD}$ at $\beta_{SU(2)} > 2$ could maintain APQCD confinement.

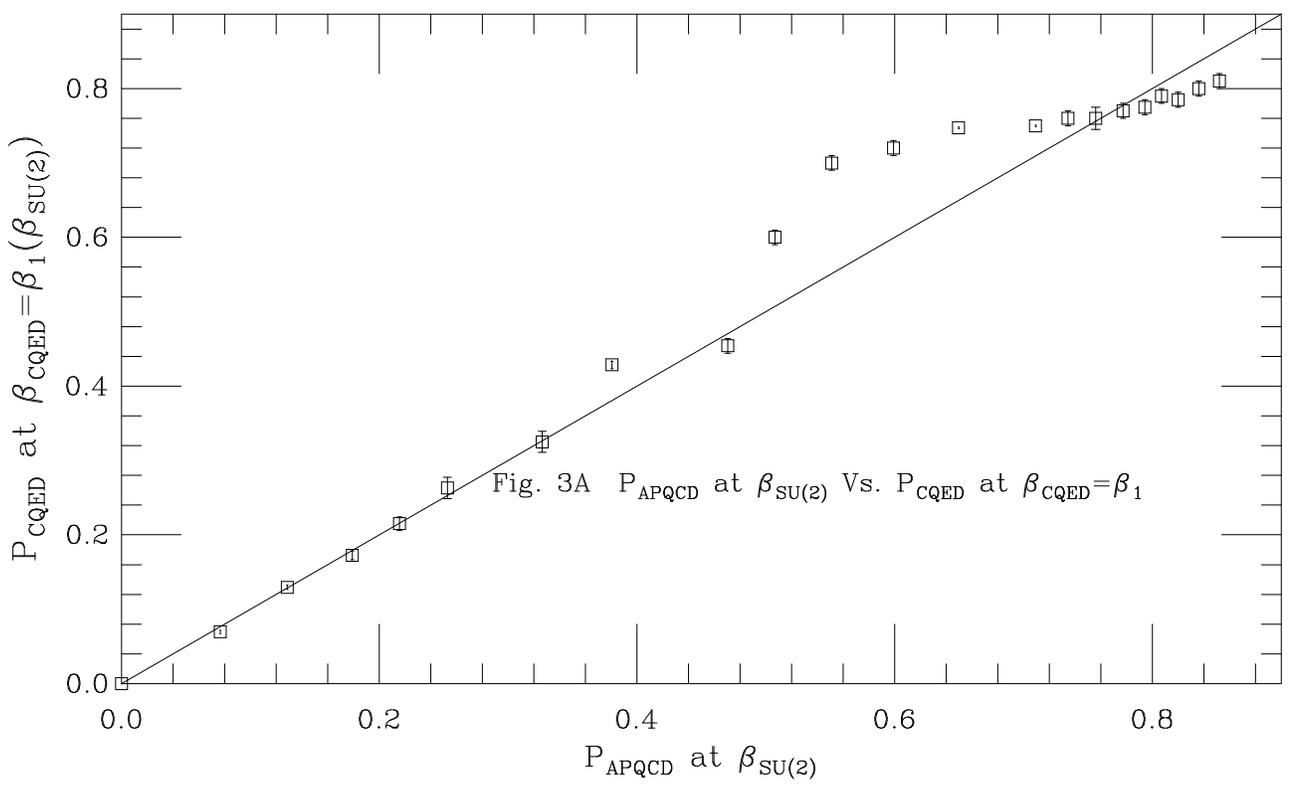

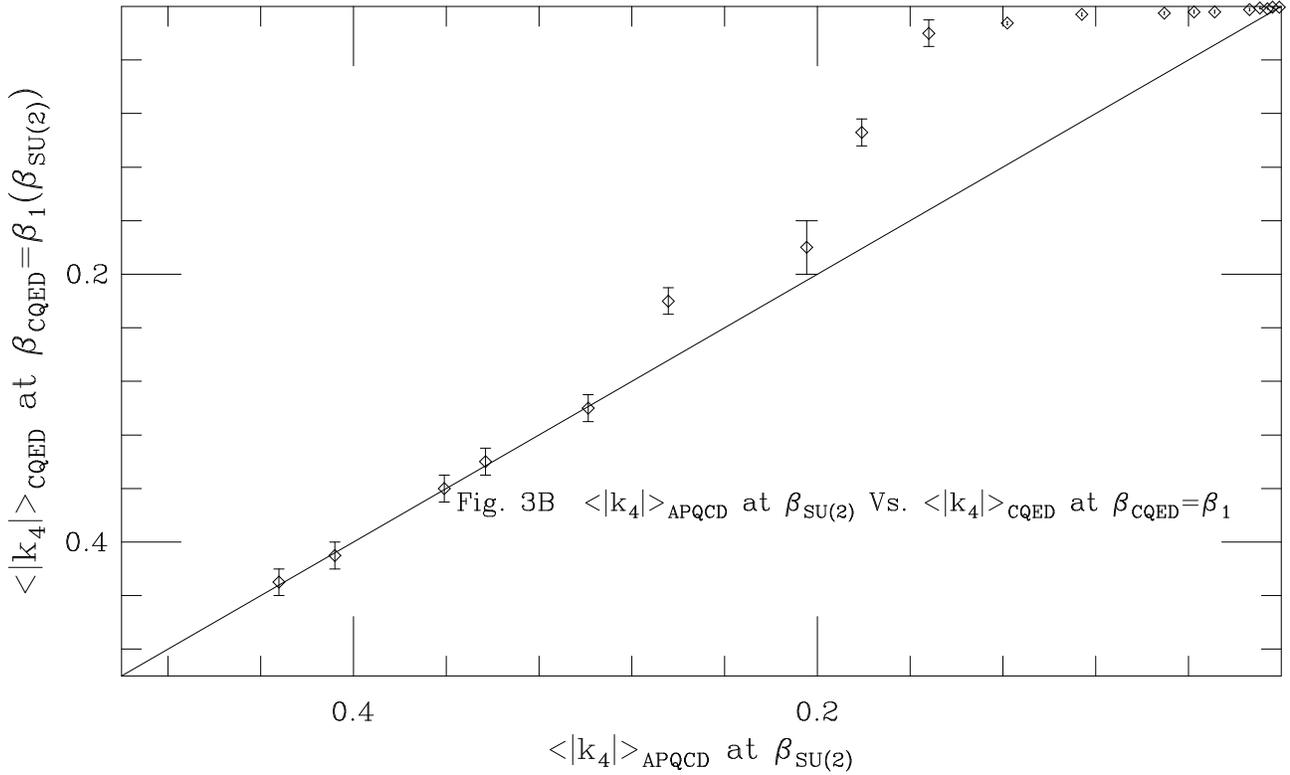

Figure 3: 3A compares APQCD plaquettes $P_{APQCD}$ at $\beta_{SU(2)}$ to CQED plaquettes $P_{CQED}$ at $\beta_{CQED} = \beta_1(\beta_{SU(2)})$ for a range of $\beta_{SU(2)}$ values. When $\beta_{SU(2)} < 2$ the data points lie on the bold $P_{CQED} = P_{APQCD}$ line showing that $S_{CQED}$ is a good model of $S_{APQCD}$. The set of points lying off of the $P_{CQED} = P_{APQCD}$ line corresponds to $\beta_{SU(2)} > 2$, when $S_{CQED}$ is not a good model of $S_{APQCD}$. 3B is an analogous plot using monopole densities.

Note that Figure 2, as characterized by Eq. (9), "explains" Abelian dominance—at least in the strong coupling regime. The $SU(2)$ plaquette in the strong coupling expansion behaves like $P_{QCD} \sim \frac{1}{4}\beta_{SU(2)}$ and the CQED plaquette like $P_{CQED} \sim \frac{1}{2}\beta_{CQED}$. Therefore, identifying $P_{CQED}(\beta_{CQED} = \beta_1)$ with $P_{APQCD}$ and applying Eq. (9) yields

$$P_{APQCD} \sim \frac{1}{4}\beta_{SU(2)} \sim P_{QCD}. \tag{11}$$

Carrying this argument over to larger Wilson loops leads to a strong coupling version[3] of Abelian dominance: at sufficiently strong coupling APQCD and QCD Wilson loop averages and, hence, string tensions are equal. Figure 4 confirms (11) and shows how this relation breaks down at weaker coupling. Interestingly, Eq. (11) contradicts the naive expectation, based on $P_{QCD}$ containing a trace over a $2 \times 2$ matrix and $P_{APQCD}$ involving no trace, that $P_{APQCD} = \frac{1}{2}P_{QCD}$.

We have obtained similar results for the $SU(3)$ Abelian projection which will be described elsewhere. For $SU(3)$, $S_{APQCD}$ is more complicated due to interspecies dynamics [11, 17, 18]. Nonetheless, we have observed completely analogous behavior in $SU(3)$. At stronger couplings, $S_{APQCD}$ is dominated by $L = q = 1$ operators; at weaker couplings, there is a crossover to a more complicated, but still $L = 1$ action. Again, $L > 1$ plaquettes are never resolved. Very preliminary $SU(3)$ results are reported in Ref. [3].

---

[2] Our demon studies, exemplified in Table 1, seem to rule out the alternative possibility that $L > 1$ plaquettes become important at larger couplings.

[3] Not to be confused with weak coupling Abelian dominance which requires only string tension equality.

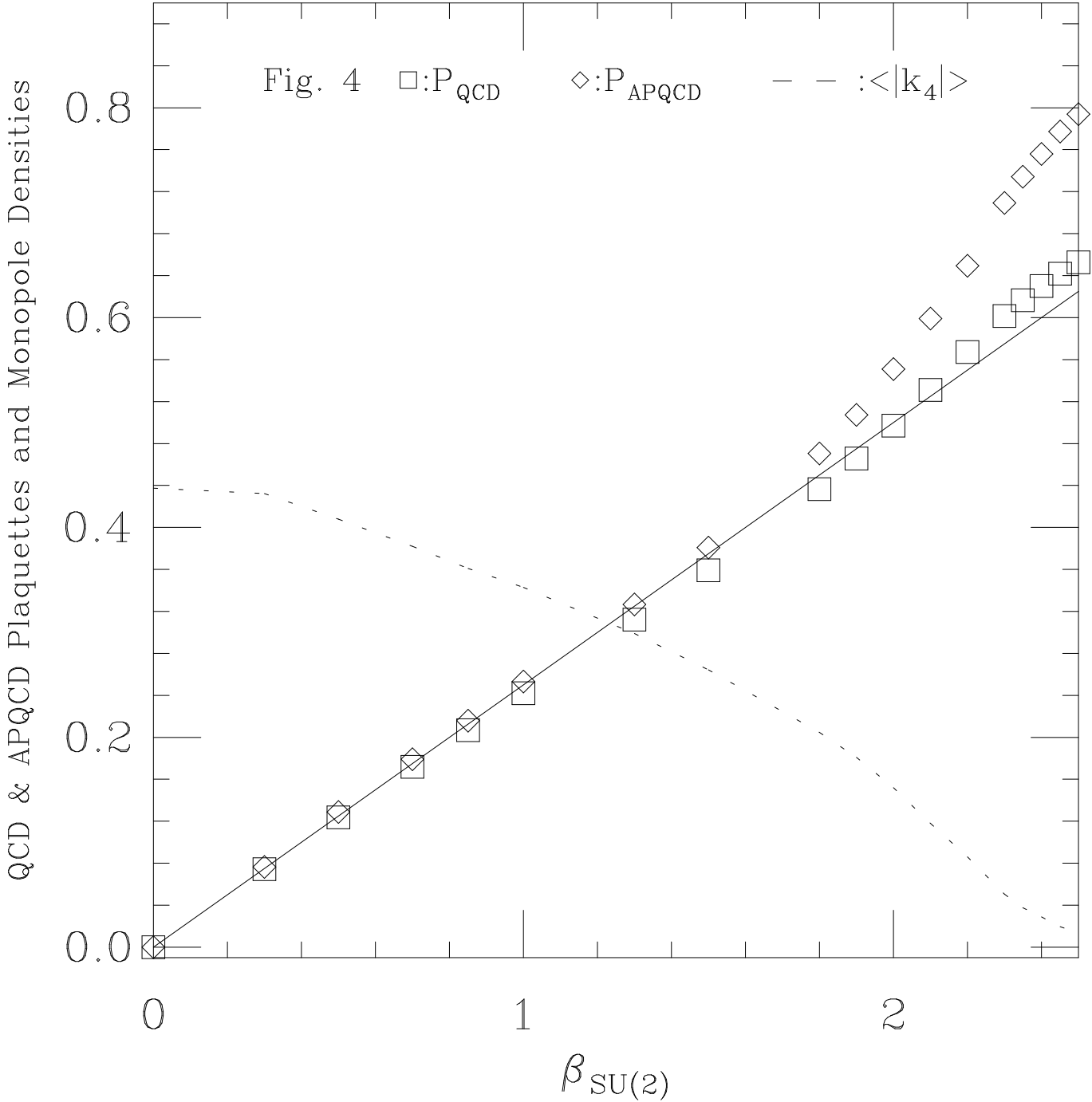

Figure 4: Figure 4 depicts the APQCD and $SU(2)$ plaquettes as a function of $\beta_{SU(2)}$. In the strong coupling region($\beta_{SU(2)} < 2$), both the APQCD and $SU(2)$ plaquettes grow like $\frac{1}{4}\beta_{SU(2)}$, the guide-to-eye line's slope. At weaker coupling($\beta_{SU(2)} > 2$) the APQCD plaquettes deviate substantially from the $SU(2)$ plaquettes. Correspondingly, the monopole density decelerates noticeably near $\beta_{SU(2)} \sim 2$.

In conclusion, while it is not form-invariant between the strong and weak coupling regimes, $S_{APQCD}$ is dominated by $L = 1$ operators for all values of $\beta_{SU(2)}$ studied. This implies that $1^3$ (rather than $L^3$) monopoles are the dominant fundamental dynamical degrees of freedom for APQCD confinement, and that phenomenological features of the APQCD confinement mechanism, such as whether APQCD is a Type I or Type II superconductor, might vary with lattice spacing. In particular, the results of this paper predict APQCD is a Type II superconductor like CQED [19] when $\beta_{SU(2)}$ is small, as distinguished from the $\beta_{SU(2)} > 2$ case studied in [20].

## 4. Acknowledgements

It is a pleasure to thank Dick Haymaker and Misha Polikarpov for stimulating discussions and MP for his $SU(2)$ FORTRAN codes. I am indebted to the Institute for Theoretical and Experimental Physics(ITEP) for their hospitality. Computing was done at the NERSC Supercomputer Center. The author is supported by DOE grant DE-FG05-91ER40617.

## Appendix A  The Demon Method

This Appendix explains the demon method [10] for ansatzes which can be link-wise partitioned so that $S^{\text{ansatz}} = \sum_{x,\mu} S^{\text{ansatz}}_{x,\mu}$ where $S^{\text{ansatz}}_{x,\mu} = \sum_{i=1}^{M} \beta_i H_i$ has the same form and coupling values $\beta_i$ on all links. $i$ labels the $M$ different operators $H_i$. While all the operators of $S^{\text{ansatz}}_{x,\mu}$ share a common link $U_{x,\mu}$, they may (and do) depend on other links. For $S^{\text{ansatz}}_B$ of Eq. (8), $M = 4$; $H_1$, $H_2$, and $H_3$ correspond to $\propto \sum_{\nu=-4}^{4} \cos q\Theta_{\mu\nu}(x)$ for $q = 1, 2, 3$; and $H_4$ is $\propto \sum_{\nu} k_{\nu}(x)k_{\nu}(x)$ summed over all directions affected by link $\theta_{\mu}^{1}$. Let

$$\Omega = \Omega(H_1, H_2, \cdots) \tag{A.1}$$

denote the number density of states in the "energy" interval $H_i$ and $H_i + \delta_i$. In statistical mechanics language, the entropy is proportional to $\log \Omega$ and, for a system with fixed total energies $E_i^T$, the inverse temperatures are

$$\beta_i \equiv \frac{\partial \log \Omega}{\partial H_i}\bigg|_{H_i = E_i^T}. \tag{A.2}$$

Now imagine a "demon" carrying $M$ thermometers corresponding to the $M$ operators $H_i$. The job of the demon is to measure the inverse temperatures $\beta_i$ of a heat bath—in our case an APQCD gauge configuration. To do this, the demon hops link-to-link and exchanges energy with the bath until it thermalizes. More precisely, at each link the demon thermometers are changed by energy increment $\Delta H_i^{\text{demon}}$ computed as follows: the bath link is randomly updated and $\Delta H_i^{\text{demon}} \equiv H_i^{\text{old}} - H_i^{\text{new}}$ is computed by evaluating $H_i^{\text{old}}$ and $H_i^{\text{new}}$ *on the bath* before and after the update. Since $\Delta H_i^{\text{bath}} = -\Delta H_i^{\text{demon}}$, the demon plus bath energy is constant under this procedure.[4] The thermometers are coupled by requiring every thermometer energy to be inside some range $[-E^0, E^0]$; if a potential update pushes any thermometer outside this range, it is rejected.

---

[4]In practice we do not retain the update of the APQCD configurations, so that the demon plus bath energy is not really constant in our procedure (as it is in Ref. [10]). Nonretention shortens the computer algorithm and avoids any possibility of damaging the APQCD configuration, a real danger since we have a whole battalion of energy-absorbing demons.

Upon denoting heat bath quantities with primes, the microcanonical partition function of the total demon-bath system is

$$\begin{aligned} Z_{mc}(E_1^T, E_2^T, \cdots) &\equiv \int [d\theta][d\theta'] \prod_{A=1}^{M} \delta(H_i + H_i' - E_T) \\ &= \int [d\theta] [dH'] \, \Omega(H_1', H_2', \cdots) \prod_{A=1}^{M} \delta(H_i + H_i' - E_i^T). \end{aligned}$$

Performing the $[dH']$ integration and Taylor expanding entropy $\log \Omega$ yields

$$Z_{mc}(E_1^T, E_2^T, \cdots) = \Omega(E_1^T, E_2^T, \cdots) \int [d\theta] \, \exp \sum_{A=1}^{M} -\beta_i H_i. \qquad (A.3)$$

Eq. (A.3) expresses the well known result that a thermalized subsystem of a microcanonical ensemble has a Boltzmann distribution with inverse temperatures given by (A.2). Hence, if we set a battalion of demons free in the importance sampling APQCD configurations, upon thermalization the demons will return with their $M$ thermometers each distributed in a Boltzmann distribution. Therefore, the $S_{x,\mu}^{\text{ansatz}}$ coupling constants $\beta_i$ corresponding to APQCD are readily extracted by fitting each of these $M$ distributions to $\exp\{-\beta_i H_i\}$.

We have tested the demon method extensively as follows. First, we generate an ensemble of $U(1)$ configurations according to a known $U(1)$ action $S_0$. For example, $S_0$ may be $S_A^{\text{ansatz}}$ at some point $P$ in $\beta_q(L)$ space. Then we apply the demon method with a trial ansatz $S_0^{\text{ansatz}}$ which, for purposes of discussion, may or may not contain all operators of true action $S_0$. If $S_0^{\text{ansatz}}$ contains all operators of $S_0$, then we find that the demon always successfully recovers $P$ and $S_0$, that is: (i)coefficients of operators in $S_0^{\text{ansatz}}$ which do not exist in $S_0$ vanish modulo statistical errors comparable in size to those in Table 1; and (ii)coefficients of operators in $S_0$ equal $P$ modulo statistical errors. The ability of the demon method to *unambiguously* and *automatically* reveal when an operator does not exist in $S_0$ is an advantage of the demon method over other methods.

If $S_0^{\text{ansatz}}$ does not contain all the operators of $S_0$, the situation is less straightforward. The demon apparently tries to obtain an effective action by using the available operators in $S_0^{\text{ansatz}}$ to fit the ensemble as optimally as possible. However, is not easy to nail down what exactly is being optimized. Therefore, it is important to simulate the ansatz action with demon couplings and verify, as we have, that ansatz expectation values reproduce APQCD expectation values.